\documentclass[12pt]{article}
\usepackage{amssymb,amsmath,cite}
\catcode`\@=11

\@addtoreset{equation}{section}
\def\theequation{\arabic{section}.\arabic{equation}}
     
\catcode`\@=11
\def\thesection{\arabic{section}.}

\def\appendix{\setcounter{section}{0}
        \def\thesection{Appendix.}
        \def\theequation{\Alph{section}.\arabic{equation}}}
\def\section{\@startsection{section}{1}{\z@}{3.5ex plus 1ex minus
   .2ex}{2.3ex plus .2ex}{\large\bf}}

\long\def\@makefntext#1{\parindent 0cm\noindent
\hbox to 1em{\hss$^{\@thefnmark}$}#1}

\begin{document}

\begin{titlepage}
\vspace{.5in}
\begin{flushright}
UCD-02-12\\
gr-qc/0209014\\
September 2002\\
\end{flushright}
\vspace{.5in}
\begin{center}
{\Large\bf
 Varying Constants, Black Holes,\\[.5em] and Quantum Gravity}\\
\vspace{.4in}
{S.~C{\sc arlip}\footnote{\it email: carlip@dirac.ucdavis.edu}\\
       {\small\it Department of Physics}\\
       {\small\it University of California}\\
       {\small\it Davis, CA 95616}\\{\small\it USA}}
\end{center}

\vspace{.5in}
\begin{center}
{\large\bf Abstract}
\end{center}
\begin{center}
\begin{minipage}{5.25in}
{\small Tentative observations and theoretical considerations  
have recently led to renewed interest in models of fundamental 
physics in which certain ``constants'' vary in time.  Assuming fixed 
black hole mass and the standard form of the Bekenstein-Hawking 
entropy, Davies, Davis and Lineweaver \cite{Davies} have argued 
that the laws of black hole thermodynamics disfavor models in 
which the fundamental electric charge $e$ changes.  I show that 
with these assumptions, similar considerations severely constrain 
``varying speed of light'' models, unless we are prepared to abandon 
cherished assumptions about quantum gravity.  Relaxation of these
assumptions permits sensible theories of quantum gravity with 
``varying constants,'' but also eliminates the thermodynamic 
constraints, though the black hole mass spectrum may still
provide some restrictions on the range of allowable models.
}
\end{minipage}
\end{center}
\end{titlepage}
\addtocounter{footnote}{-1}

\section{Introduction}

The idea that the fundamental ``constants'' of our Universe may vary in 
time dates back at least to Dirac's Large Number Hypothesis \cite{Dirac}.  
Until recently, physically interesting variations seemed to be excluded 
by observation.  Over the past several years, however, Webb et al.\ 
\cite{Webb,Murphy} have reported evidence that the fine structure constant 
$\alpha$ may have been slightly smaller in the early Universe.  While 
this claim is still far from being established, the possibility, along with 
work on cosmological implications of ``varying constants,'' has inspired 
renewed interest in models in which either the elementary charge $e$ or 
the speed of light $c$ is dynamical; for a sample of this work, see 
\cite{Bek0,Moffat,Albrecht,Clayton,Barrow,Sandvik,Magueijo,%
Bassett,Chako,Calmet,Olive}.

In a Brief Communication to {\it Nature} \cite{Davies}, Davies, Davis 
and Lineweaver contend that black hole thermodynamics favors
models with a varying speed of light.  Their basic argument is simple.
The Bekenstein-Hawking entropy of a charged black hole of mass $M$ 
and charge $Q$ is
\begin{equation}
S/k = \frac{\pi G}{\hbar c}\left[ M + \sqrt{M^2 - Q^2/G}\right]^2 .
\label{a0}
\end{equation}
Suppose $\alpha$ is indeed increasing in time, as  Webb et al.\ suggest.  
If this  variation comes from an increase in $e$, the resulting increase in 
$Q$ will cause the entropy of such a black hole to decrease, apparently violating 
the generalized second law of thermodynamics. If, on the other hand, the 
change comes from a decrease in $c$, the entropy will increase with time, as 
it should. 
 
This is an intriguing argument, but it requires several key assumptions:
\begin{enumerate}
\item the Bekenstein-Hawking formula, and in particular (\ref{a0}),
remains a good approximation for black hole entropy in a theory with 
``varying constants'';
\item Planck's constant $\hbar$ and Newton's constant $G$ remain
constant;
\item it is sufficient to look at the entropy of the black hole alone, and
not its environment;
\item the black hole mass $M$ remains constant as $\alpha$ varies.
\end{enumerate}
While these assumptions seem plausible, they need not be correct.
Indeed, there are particular models in which each is violated 
\cite{Magueijo2,CarVai}, and a full analysis would require a much
more specific and detailed theory.  Still, these assumptions offer an 
interesting ``phenomenological'' starting point for investigating the 
broader question of whether, and to what extent, black hole quantum 
mechanics can constrain theories of ``varying constants.''

In the first part of this paper, I  show that the implications of assumptions 
1--4 are far more radical than Davies et al.\ suggest, and lead to predictions 
unpalatable enough to militate against ``varying speed of light'' models.  
I then discuss the options that become available when one relaxes
these assumptions.  The resulting loopholes are wide enough to allow 
sensible quantum theories of gravity with ``varying constants''---string 
theories with time-dependent compactification radii are useful examples%
---though, of course, the thermodynamic constraints of Ref.\ \cite{Davies} 
are then lost.  Still, while black hole quantum mechanics will not constrain
{\it all\/} models of ``varying constants,'' it may still narrow the range of
models we must consider.

\section{Varying constants and the black hole mass spectrum}

Consider a Reissner-Nordstrom black hole with charge $Q=qe$ and 
mass $M=\sqrt{r}M_{P}$, where $M_{P} =  (\hbar c/G)^{1/2}$ is the 
Planck mass.  Quantization of charge requires that $q$ be an integer.  
In simple models of black hole thermodynamics (see \cite{Bek} for a
review), $r$ is an integer, or a fixed constant multiple of an integer, as well.  
More elaborate approaches to quantum gravity lead to more complicated 
black hole spectra: for instance, neutral black holes in loop quantum 
gravity have \cite{loop}
\begin{equation}
r = \frac{\gamma}{4}\sum_i\sqrt{p_i(p_i+2)}
\label{b1}
\end{equation}
where the $p_i$ are arbitrary integers and $\gamma$ is a constant of order 
unity, while the string theoretical black holes of Ref.\ \cite{string} depend 
on four pairs of integers $(p_i,{\bar p}_i )$ that count branes wrapped around 
various compactified dimensions, with
\begin{equation}
r = \gamma'\left( \frac{p_1{\bar p}_1p_2{\bar p}_2p_3{\bar p}_3}{p_4{\bar p}_4}
   \right)^{1/2}\left(\sum_{i=1}^4 \left[
   \sqrt{\frac{p_i}{{\bar p}_i}} + \sqrt{\frac{{\bar p}_i}{ p_i}}\right]\right)^2 .
\label{b2}
\end{equation}
The minisuperspace model of Barvinsky et al.\ \cite{Barvinsky} has
\begin{equation}
r = \frac{1}{4}\frac{(2n+1)^2}{2n+1+  q^2\alpha} +  q^2\alpha ,
\label{b2a}
\end{equation}
where $n$ and $q$ are integers, and, as above, $Q=qe$.
In each of these examples, though, and in virtually all other models that 
have been considered, $r$ is still discrete.  Indeed, it is hard to see how 
to reconcile a finite Bekenstein-Hawking entropy with a continuous black 
hole mass spectrum.  

Given the quantum numbers $q$ and $r$, the entropy (\ref{a0}) 
becomes 
\begin{equation}
S/k = \pi\left[\sqrt{r} + \sqrt{r  -  q{}^2\alpha}\right]^2 
\label{a1}
\end{equation}
(see also \cite{Duff}).  Clearly, an increase in $\alpha$ will lead to a 
decrease in the entropy, and an apparent violation of the generalized second 
law of thermodynamics, unless $q$ or $r$ also evolve.  But if $q$ and $r$ 
are discrete, they can change only in finite jumps (except, of course, through
changes in parameters like $\gamma$ in (\ref{b1}) or $\gamma'$ in (\ref{b2}),
which may evolve continuously in models like string theory with time-dependent
moduli).  This quantization suggests a discrete evolution for $\alpha$ as well, 
an ingredient not easily incorporated in many current models.  But even 
without such a feature, the discrete nature of $r$ presents a serious problem 
for a number of models with varying $c$, one already evident for uncharged 
black holes.

Following Davies et al., I assume for now that $M$ remains constant. 
The speed of light then enters $r$ through the Planck mass.\footnote{Duff 
\cite{Duff} has criticized models involving variations of dimensionful 
parameters.  I take ``varying $c$'' as shorthand for ``variation of all 
dimensionless parameters, such as $m_e/M_{P}$, that depend on $c$.''  
The disentangling of dimensionful parameters is still somewhat ambiguous, 
but given a standard choice of conventions, this prescription leads to a 
well-defined model.}   It is then easy to see that given a black hole with 
``mass quantum number'' $r$, a change $\Delta c$ requires a jump
\begin{equation}
\Delta r  = (\Delta x)r \quad \hbox{with}\
\Delta x = -\frac{\Delta c}{c} \left( 1 + \frac{\Delta c}{c} \right)^{-1} .
\label{a2}
\end{equation}
Note that a positive $\Delta x$ corresponds to a decrease in $c$, and an 
increase in $\alpha$.  In some models, Planck's constant $\hbar$ also 
varies with $c$ \cite{Magueijo}.  Such a variation would lead to a few minor 
changes---$r$  and $\Delta r$ in (\ref{a2}) would  be replaces by $r^\beta$ 
and $\Delta(r^\beta)$ for some exponent $\beta$---but the qualitative 
considerations below would not be affected.  (Coule \cite{Coule} has 
discussed related problems in cosmology coming from such a variation 
of $M_P$, and Banks et al.\ \cite{Banks} have pointed out serious fine 
tuning issues.)

The first thing to notice about eqn.\ (\ref{a2}) is that for a given $r$ or a given 
$\Delta c$, it may not have any solutions.  That is, suppose we are given a quantum
theory of gravity that determines a spectrum $\cal S$ of possible values for $r$.  
Then for any particular $r\in{\cal S}$ and any fixed $\Delta c$, $r+\Delta r = 
(1 + \Delta x)r$ may not lie in $\cal S$.  This situation can have two interpretations, 
which I will expand upon below: either many black hole masses are forbidden, even 
for allowed quantum numbers $r\in{\cal S}$, or else the allowed changes $\Delta c$ 
are sharply constrained by the presence of black holes.

For illustration, let us begin with a model in which the allowed values of $r$, 
and thus $\Delta r$, are integers.  For eqn.\ (\ref{a2}) to have any solutions, 
${\Delta x}$ must be rational: ${\Delta x} = p/N$ and $r=nN$, where $p$, $n$, 
and $N$ are integers, with $p$ and $N$ relatively prime.  It follows that 
$\Delta c/c = -p/(p+N)$.  But the total variation ${\Delta\alpha/\alpha}$  
claimed by Webb et al.\ is only on the order of $10^{-5}$, and larger variations 
in $\alpha$ are excluded by other observations \cite{alpha}, so ${\Delta c/c}$ 
should be no  larger than about $10^{-5}$ at each discrete step.   We must thus 
require $N\gtrsim10^5$.  

The model thus forbids black holes with masses less than some  minimum 
value $M_0 = \sqrt{N}M_P\gtrsim350\,M_{P}$, and it requires that {\it all\/} 
black hole masses be multiples of $M_0$ by the square root of an integer.    Note 
that this estimate of $M_0$ is conservative: I have assumed not only that 
the change in $\alpha$ is as large as that reported by Webb et al., but also that 
it all came in a single jump.  It is easy to see that if $\Delta c$ is spread over $k$ 
equal steps, $M_0$ increases by a factor of $\sqrt{k}$.  If, for instance, jumps  
occur at a characteristic electromagnetic time scale $\tau = \hbar/m_ec^2$, 
one finds $M_0\sim 10^{16}\,g$, excluding standard primordial black holes.

The existence of a change $\Delta c$ thus restricts the allowed black hole masses.
Conversely, the existence of black holes limits the possible changes in $c$: given a 
collection of black holes with mass quantum numbers $\{r_i\}$, ${\Delta c/c}$
is restricted to be of the form $-p/(p+N)$, where $p$ and $N$ are relatively prime
and $N$ is a common divisor of the $r_i$.  If the greatest common divisor of the 
$r_i$ is less than about $10^5$, no variation in $c$ is compatible with observation.  
Note also that the restriction on $\Delta c/c$ is time-dependent: a jump in $c$ 
causes a corresponding shift $r_i\rightarrow (1 + \Delta x)r_i$, leading to a new 
condition on any future jump in $c$.

Integral quantization of $r$ may well be too simple, but the model 
demonstrates the key features present for more complicated spectra.
Eqn.\ (\ref{a2}) can be satisfied in two ways:
\begin{enumerate}
\item in any region containing a black hole of mass $\sqrt{r}M_{P}$, allow 
only a sharply limited set of changes in $c$ (choose $\Delta c/c$ as a function 
of $r$ so that  $r+\Delta r$ is in the spectrum); or
\item for fixed ${\Delta c/c}$, allow only a sharply limited set of black holes  
(restrict $r$ to those values for which both $r$ and $(1 + \Delta x)r$ 
are in the spectrum).
\end{enumerate}
Alternative 1 requires a mysterious new local coupling between $c(t)$ and black 
hole mass.  One could imagine treating the restriction as a boundary value 
problem for $c(x,t)$, though this would require boundary conditions that evolve
as black hole masses change: the global evolution of $c(t)$ would be affected
each time a black hole captured an electron or emitted a quantum of Hawking 
radiation,  and boundary conditions would multiply as new black holes form 
and vanish as old black holes evaporate.  It is not obvious that such boundary 
conditions are consistent, but if they are, a new problem will appear---if $c$ is 
permitted to jump by different discrete steps in different regions, it becomes 
difficult to maintain the observed large-scale homogeneity of the fundamental 
constants.

Alternative 2 drastically reduces the number of permitted states of a black 
hole.  The allowed states will typically be no less sparse than they were in 
our illustrative example.  For models such as loop quantum gravity, mass
spacings can become very small at high masses \cite{Barreira}, and eqn.\
(\ref{a2}) will have many {\it approximate\/} solutions, but only for very
peculiar spectra will many black hole masses give {\it exact\/} solutions
(see below).  Note that the excluded states occur in all mass ranges; this 
is not merely a Planck scale effect that can be blamed on our ignorance 
of quantum gravity.  Apart from the difficulty of implementing such a 
restriction in existing models, such a reduction in the number of allowed 
states would seem to invalidate any statistical mechanical interpretation 
of the Bekenstein-Hawking entropy as a measure of the number of 
accessible microscopic states.  

With either alternative, one is likely to have a problem with ``small'' 
black holes, those large enough that semiclassical approximations should be
reliable but small enough that the selection rule (\ref{a2}) is highly restrictive.
For the case of integer spacing, we saw that there was a large mass gap between 
the Planck scale and the first allowed black hole.  Other spectra may not have a 
completely empty gap, but for spectra depending on more than one integer, 
mass spacings are generally much wider at low masses than at high masses 
\cite{Barreira}, so one expects a paucity of admissible low-mass black holes.

\section{Searching for loopholes}

At a minimum, these considerations severely constrain ``varying speed 
of light'' models that satisfy assumptions 1--4 of the introduction.  It may 
well be that the observations of Webb et al.\ are wrong, and that $\alpha$ 
is actually constant.  But until the observational situation is more settled, it 
is worth investigating other ways out of this dilemma.

One possibility is to give up quantization of black hole masses.  This
is not a very comfortable choice: it would require not only that we abandon 
most existing approaches to quantum gravity (loop quantum gravity, 
most string theoretical models), but that we give up powerful theoretical ideas 
like holography \cite{Susskind} that depend on the finiteness of the number 
of black hole states.  Of course, it could be that black holes in ``varying speed
of light'' theories are drastically different from those in standard general
relativity \cite{Magueijo2}, in which case these issues would have to be 
rethought.  But this would require that we discard even the little we believe
we now understand about black hole entropy and quantum gravity.

Another possibility is to look for a spectrum $\cal S$ of black hole states for 
which the constraints we have found are unimportant.  Such a spectrum
exists, but it is physically unrealistic: it requires that $\log{M/M_P}$ be 
integrally spaced.  To see this, let $\Delta c/c =-\Delta x/(1+\Delta x)$ be 
an allowed change in $c$ at time $t_0$, and suppose that a black hole with 
mass quantum number $r_0$ is present.  This requires that both $r_0$ and 
$(1+\Delta x)r_0$ be in $\cal S$.  If we now require that  black holes with 
$r=(1+\Delta x)r_0$ also be allowed at time $t_0$, $(1+\Delta x)^2r_0$ must 
also be in $\cal S$.  Continuing this argument, one is lead to a spectrum 
$2\log(M/M_P) = \log{r_0} + k\log(1+\Delta x)$, where all integers $k$ must 
occur.  Such a spectrum is increasingly sparse at high masses, and would lead 
to rather odd predictions.  For example, if $\Delta x \sim 10^{-5}$, one would 
not be able to drop the Earth into a Solar-mass black hole: no mass states would 
be available.  Even if jumps in $c$ occur at  an electromagnetic time scale $\tau 
= \hbar/m_ec^2$, so $\Delta x \sim 10^{-42}$, the spectrum of supermassive 
black holes would still have a spacing of several grams: one couldn't drop a dime 
into the black hole at the center of the Milky Way. 

A third possibility is that while $c$ varies cosmologically, it remains constant
at black hole horizons \cite{Barrow2}.  This is the case for single static black 
hole solutions in certain models of ``varying constants,''  essentially as a
result of no hair theorems.  It is not at all clear, however, that this argument 
can be extended to dynamical solutions with more than one black hole; such 
boundary conditions are strong enough that there may be no solutions except
those with globally constant $c$.   The process of new black hole formation
seems especially problematic.  While one could imagine a process in which
a collapsing star radiates away varying $c$ ``hair,'' it is hard to see why
$c$ should freeze to the same value at widely separated black holes formed
at very different times; but without such constancy, it is even harder to 
understand the observed spatial homogeneity of fundamental constants.

\section{Modifying the initial assumptions}

The final possibility is to abandon one of the four assumptions enumerated
in the introduction.  This would, of course, invalidate the original argument 
of Ref.\ \cite{Davies} that black hole thermodynamics constrains varying
constants.  But one may ask whether quantization of the black hole mass 
spectrum continues to provide any useful constraints.

The first assumption of Ref.\ \cite{Davies} was that the Bekenstein-Hawking 
formula remains a good approximation for the entropy in theories with ``varying 
constants.''  Such theories necessarily contain at least one new dynamical field, 
$\alpha$ itself, which could contribute to the  entropy.  We know that $\alpha$ 
varies very slowly, if at all, in time, so the standard Bekenstein-Hawking entropy 
is plausibly a good approximation.  But $\alpha$ may vary significantly in 
{\em space\/} near a black hole horizon; indeed, in some models \cite{Magueijo2}, 
$\alpha$ goes to zero or infinity at the horizon of an isolated static black hole.  
In such a situation, the thermodynamic argument of Davies et al.\ clearly fails, 
and black hole thermodynamics imposes no obvious restrictions on ``varying 
constants.''   The black hole mass spectrum is a different matter, though; as 
long as black hole solutions exist, quantization of this spectrum will continue 
to constrain models, independent of purely thermodynamic considerations.

The second starting assumption was that $\hbar$ and $G$ remain constant
as $\alpha$ changes.  One can, of course, postulate ad hoc changes in these 
parameters in such a way as to save the second law of thermodynamics, although 
it is not clear how to justify such an assumption from first principles.  Again,
however, as long as black hole masses are quantized and the Planck mass
evolves, the nonthermodynamic constraints on admissible models described
above will remain.

The third assumption was that it is sufficient to look at the entropy of a black 
hole alone, and not its environment.  This is probably incorrect; it is argued 
elsewhere \cite{CarVai} that the change in the Hawking temperature coming from 
a variation in $\alpha$ induces heat flow, which in turn affects the black hole 
equilibrium mass, fundamentally altering the entropy balance.  Once again,  
though, quantization of the black hole mass spectrum continues to constrain 
models, independent of thermodynamic considerations.

The fourth assumption was that the mass $M$ of a black hole remains 
constant as $\alpha$ varies.  Relaxing this assumption probably provides
the most important loophole.  For example, in the minisuperspace quantization 
of Barvinsky et al.\ \cite{Barvinsky}, the black hole mass $M$ depends explicitly 
on the fine structure constant, and it may be checked from eqns.\ (\ref{b2a}) and 
(\ref{a1}) that the $\alpha$ dependence of the entropy disappears.  The cause 
of this ``miraculous'' cancellation is easy to understand: by construction, the 
fundamental quantum observables in this model are the charge quantum 
number and the entropy, while black hole mass is a secondary, derived 
quantity.

A more compelling example comes from string theory.  In the string theoretical 
models of Ref.\ \cite{string}, the fine structure constant depends on various radii 
of compactification, and can change as those radii evolve.  Nevertheless, the 
entropy is independent of the compactification radii; in the notation of eqn.\ 
(\ref{b2}), it is
\begin{equation}
S/k = 2\pi \prod_{i=1}^4 \left( \sqrt{p_i} + \sqrt{{\bar p}_i} \right) ,
\label{d1}
\end{equation}
and depends only on the integers $\{p_i,{\bar p}_i\}$ that count the number
of branes.  If one reexpresses the entropy (\ref{d1}) in terms of mass and charges,
the  usual Bekenstein-Hawking formula can be recovered, but the mass and charges 
again depend on the compactification radii in just the right way to ensure that 
these radii cancel from the entropy.  

Again, the physics lying behind this cancellation is not hard to understand.  
Near-extremal black holes in string theory can be thought of as comprising a 
collection of weakly coupled branes.  These branes are wrapped around compactified
spatial dimensions, and their masses depend on the sizes of these dimensions---%
that is, on the same compactification radii that determine the fine structure 
constant.  The masses and charges therefore depend on moduli such as 
compactification radii, and can vary if these radii vary.  The entropy (\ref{d1}) 
and the mass quantum number (\ref{b2}), on the other hand, are fixed by 
the numbers of branes, and thus decouple from any changes in the moduli 
\cite{string,HMS}.

A related argument has been advanced by Flambaum \cite{Flambaum} in 
the context of ``phenomenological'' models in which black hole areas are 
integrally quantized.  The Bekenstein-Hawking entropy (\ref{a0}) can be rewritten 
suggestively as
\begin{equation}
S/k = \frac{\pi c^3}{\hbar G}r_+{}^2 ,\qquad
M = \frac{r_+c^2}{2G} + \frac{Q^2}{2r_+c^2}
\label{a3}
\end{equation}
where the event horizon is located at $r=r_+$.  In this formulation, it seems natural to 
guess that a change in $\alpha$ affects the ``electrostatic self-energy'' contribution
to the black hole mass while leaving $r_+$, and thus the entropy, fixed.  As Dicke
noted long ago \cite{Dicke}, however, such a variation of electrostatic energy  in 
ordinary matter could lead to violations of the weak equivalence principle.\footnote{%
Bekenstein has recently shown that for a certain class of models with 
varying $e$, a scalar coupling modifies Coulomb's law in a manner that compensates 
for position dependence of electrostatic self-energy \cite{Bekenstein2}, but in such 
models it seems rather unlikely that eqn.\ (\ref{a0}) will continue to hold even 
semiclassically.}   Current limits on such variation are comparable to the reported 
observations of Webb et al., and planned experiments should give an improvement 
of several orders of magnitude \cite{alpha}, offering at least an indirect experimental 
test for such a picture.

\section{Conclusions}

The first conclusion of this analysis is that it can be risky to speculate about
the effects of ``varying constants'' without a concrete model.  We have seen that
a simple set of plausible assumptions leads to drastic and implausible conclusions 
about the quantum mechanics of black holes, but also that these assumptions 
are violated in particular models, including, notably, string theory.  In particular,
this makes the thermodynamic arguments of Ref.\ \cite{Davies}, which are based
on these assumptions, suspect.

Second, though, we have found a new set of criteria that continue to place some 
constraints on models of ``varying constants'' even when the purely thermodynamic 
analysis fails \cite{CarVai}.  The arguments based on black hole quantum mechanics 
certainly do not rule out models with varying $e$ or $c$---as I have stressed,
they depend on assumptions about black hole masses that do not hold in models
such as those coming from string theory.  But they limit the spectrum of allowable 
models, and also provide a simple way to screen out some ``phenomenological'' 
descriptions that are not based on  a detailed theoretical framework.

In particular, suppose the recent claims of observable changes in the fine 
structure constant are confirmed, for example by precision measurements of 
the cosmic microwave background \cite{Martins}.  Such an event would confront 
existing models of ``varying constants'' with a new challenge, demanding detailed, 
testable predictions of time evolution.  But such a radical departure from standard 
physics would also call upon us to explore a wide range of new, and quite possibly 
incomplete, ideas.  Under those circumstances, the requirements of black hole 
quantum mechanics---and, in particular, of a sensible mass spectrum---could 
provide useful constraints on the space of theories to be investigated.

\vspace{1.5ex}
\begin{flushleft}
\large\bf Acknowledgments
\end{flushleft}

I would like to thank David Coule for pointing out an error in the first version
of this paper.  This work was supported in part by U.S.\ Department of Energy 
grant DE-FG03-91ER40674.

\end{document}